\documentclass{article}
\usepackage{arxiv}

\usepackage[utf8]{inputenc} 
\usepackage[T1]{fontenc}    
\usepackage{hyperref}       
\usepackage{url}            
\usepackage{booktabs}       
\usepackage{amsfonts}       
\usepackage{nicefrac}       
\usepackage{microtype}      
\usepackage{lipsum}		
\usepackage{graphicx}
\usepackage{doi}
\usepackage{subcaption}
\usepackage[numbers]{natbib}

\title{A Multi-factorial Analysis of Polarization on Social Media}

\author{ \href{https://orcid.org/0000-0003-1634-8856}{Célina TREUILLIER}\\
	Université de Lorraine, CNRS, LORIA\\
	Nancy, FRANCE\\
	\texttt{celina.treuillier@loria.fr} \\
	\And
	\href{https://orcid.org/0000-0003-4252-4526}{Sylvain CASTAGNOS} \\
	Université de Lorraine, CNRS, LORIA\\
	Nancy, FRANCE\\
	\texttt{sylvain.castagnos@loria.fr} \\
 	\And
	\href{https://orcid.org/0000-0002-9876-6906}{Armelle BRUN} \\
	Université de Lorraine, CNRS, LORIA\\
	Nancy, FRANCE\\
	\texttt{armelle.brun@loria.fr} \\
}

\date{}

\renewcommand{\headeright}{}
\renewcommand{\undertitle}{}

\begin{document}
\maketitle

\begin{abstract}
Polarization is an increasingly worrying phenomenon within social media. Recent work has made it possible to detect and even quantify polarization. Nevertheless, the few existing metrics, although defined in a continuous space, often lead to a unimodal distribution of data once applied to users' interactions, making the distinction between polarized and non-polarized users difficult to draw. 
Furthermore, each metric relies on a single factor and does not reflect the overall user behavior. Modeling polarization in a single form runs the risk of obscuring inter-individual differences. In this paper, we propose to have a deeper look at polarized online behaviors and to compare individual metrics. 
We collected about 300K retweets from 1K French users between January and July 2022 on Twitter. Each retweet is related to the highly controversial vaccine debate. Results show that a multi-factorial analysis leads to the identification of distinct and potentially explainable behavioral classes. This finer understanding of behaviors is an essential step to adapt news recommendation strategies so that no user gets locked into an echo chamber or filter bubble. 
\end{abstract}

\section{Introduction}
\label{intro}
The influence of social media (SM) is tremendously growing worldwide. A recent study from the Pew Research Center~ \footnote{https://www.pewresearch.org/journalism/2020/07/30/americans-who-mainly-get-their-news-on-social-media-are-less-engaged-less-knowledgeable/} has shown that one in five adults gets her news primarily through SM, and tends paradoxically to be less well-informed. The impact of SM on polarization can then be explained in two ways. On the one hand, a lack of information due to filter bubbles and echo chambers can affect the degree of polarization of users~\cite{azzimonti2022polarization}. On the other hand and paradoxically, the more a polarized user is exposed to opposing views on SM, the more her degree of polarization increases~\cite{bail2018opposing}. The balance to be found in terms of diversity of opinions, sources, and content is therefore extremely delicate.

Recent studies suggest that it would be more appropriate to tailor the level of diversity and recommendation strategies to behavioral classes ~\cite{bernstein2020diversity, stray2021designingrs, treuillier2022being} rather than to maximize diversity in the same way for all users~\cite{helberger2018diversity, lunardi2020filter, heitz2022benefits}. 
In order to do so, user polarization behaviors need to be understood and modeled in detail. Our work is a first step towards this goal of adapting the recommendations.

Polarization has been investigated in the literature from two different perspectives: a network perspective which mainly relies on the SM structure in order to quantify the polarization of a community and to highlight the content of discussions~\cite{badami2017peekingit}, and an individual perspective to measure the polarization and the impact of recommender systems through the diversity level of accepted recommendations~\cite{moller2018not}. In this paper, we focus on this second perspective, since we aim to quantify the level of polarization at the individual level, and we define 2 research questions. \textbf{RQ1}: Do the current individual polarization metrics contribute to distinguishing polarized users from non-polarized users? \textbf{RQ2}: Can a multi-factorial analysis identify different classes of polarization behavior?

In the rest of this paper, section~\ref{SOA} consists of a literature review about polarization metrics on SM. Section~\ref{caseStudy} details the experimental setup. The multi-factorial analysis is presented in Section~\ref{multi_analysis}, while conclusions and perspectives are drawn in Section~\ref{CCL}.

\section{Literature Review}
\label{SOA}

SM play an important role in polarization~\cite{kubin2021role,vanbavel2021media}. Conover~\textit{et al.} were the first to study polarization on Twitter~\cite{conover2011political}. Since then, many polarization metrics have been put forward and used in the literature, such as modularity~\cite{newman2006modularity}, controversy~\cite{garimella2018quantifying}, or metrics based on a probability density function~\cite{morales2015measuring}. 
Though these polarization metrics rely on a varied set of information, they share a characteristic: they quantify overall polarization using a graph representing users and their interactions. However, as well as being explained by social or technological filters, polarization is also influenced by individual factors~\cite{geschke2019triple} and only a few metrics are designed to quantify individual polarization. 

Among these individual metrics, we can find the \textit{polarization score} proposed by Becatti~\textit{et al.}~\cite{becatti2019extracting}. The latter is based on the identification of a set of communities $C$. The polarization score of a user $u$ depends on the ratio of interactions in each community, with $N_{u,c}$  the number of interactions of $u$ in community $c$, and $N_u$ her total number of interactions. The polarization score $\rho(u)$ of user $u$ corresponds to the maximum of this ratio, as presented in Equation~(\ref{becatti_metric}).

\begin{equation}
\label{becatti_metric}
    \rho(u)= \max_{c \in C}\big\{\frac{N_{u,c}}{N_u}\big\}
\end{equation}

Highly polarized users, accessing a unique community have $\rho = 1$, while users equally accessing all the communities have $ \rho = 1/C$. In our view, this metric has several limits: its minimum bound depends on the number of communities, it only takes into account the community with which the user has interacted the most, and it does not inform about which communities are accessed. In a two-community context, the polarization score presented by Schmidt~\textit{et al.} is similar but ranges between $-1$ and $1$ and is oriented: the value informs about which community is accessed more often~\cite{schmidt2018polarization}.

To go further, Cicchini~\textit{et al.}~\cite{cicchini2022news} propose the {\it Lack of Diversity} (LD) metric that is, to some extent, highly similar to the \textit{polarization score} of Becatti~\textit{et al.}~\cite{becatti2019extracting}. The main difference lies in the fact that it considers sources of information a user interacted with, concretely a set of $M$ media outlets, rather than communities. Each user $u$ is represented by her number of interactions $N_{u,m}$ on news from media $m$. LD is computed as follows: 
\begin{equation}
\label{cicchini_metric}
    LD(u) = \max_{m \in M}\big\{N_{u,m} \cdot log(\frac{|U|}{|U_m|})\big\} 
\end{equation}

$|U|$ is the total number of users and $|U_m|$ is the number of users interacting with media $m$. The term $log(\frac{U}{U_m})$ corrects a potential bias introduced by $m$ when shared by a large number of users. As in~\cite{becatti2019extracting}, $LD$ represents the maximum value within the vector. Thus calculated, the LD metric is not bounded, and should therefore be normalized. 

These graph-based metrics, which quantify polarization at the individual level, face a main limit: they are only computed on a single factor (communities or media outlets). However, polarization is known to occur over the influence of multiple factors~\cite{chen2019modeling, sunstein1999law}. As a consequence of considering only a single factor, several users may be identified as similarly polarized, but may in fact exhibit a wide range of behaviors and distinguish themselves in different ways. 

To summarize, while different metrics have been proposed in the literature, individual polarization metrics are still scarce. Besides, to the extent that they are based on each user's preferred behavior (\textit{i.e.}, the maximum value of the observed variable), we question their ability to accurately differentiate polarization behaviors on SM. This is why we propose to gradually conduct a multi-factorial analysis.

\section{Experimental Setup: Polarization about the Vaccine Debate on Twitter}
\label{caseStudy}

To answer our research questions (\textbf{RQ1}) and (\textbf{RQ2}), we propose to study a real SM context. We focus on the highly controversial vaccine debate, which was widely discussed following the COVID-19 crisis. 

\subsection{Data Collection}
\label{dataCollection}

We used the Twitter API (\textit{v2}), with academic research access. Data collection relies on the concept of elite users~\cite{primario2017measuring} that represent users who are relevant to the subject matter. 
We assume that elite users' tweets, when related to the selected topic, are always in line with their beliefs. Inspired by the methodology of Primario~\textit{et al.}, we fix conditions that elite users must satisfy to ensure their legitimacy:
(1) have a significant number of followers;
    (2) personally manage their Twitter account;
    (3) are known by the general audience, through media or government interventions;
    (4) are qualified by education and/or profession to address the subject of matter.  

Elite users are an effective entry point for collecting data about a specific topic because their opinions are publicly known~\cite{primario2017measuring}. Nevertheless, as our objective is to analyze standard users' interaction behaviors, it is necessary to have a faithful overview of standard users' interactions about the selected topic during a specific period. The dataset must be balanced in terms of opinion carriers, and representative of behaviors adopted on SM about a specific topic. 

To obtain such a dataset, we carried out several steps, run after having chosen the topic, identified a relevant set of elite users, and defined a collection period. These steps are: (1) Collect all tweets published by the set of elite users during the predefined period; (2) Filter tweets about the topic of interest; (3) Collect information about a random subset of interacting standard users for each collected tweet; (4) Identify the most active standard users among those selected in Step~3; (5) Collect all interactions of selected standard users on collected elite users' tweets during the defined period. 

Following the procedure detailed above, we manually identified 20 French-speaking elite users having a legitimate voice in the vaccine debate (10 pro-vaccine and 10 anti-vaccine). Their opinion is known because they have clearly expressed it publicly, and the community to which they relate is therefore unambiguous. To preserve their confidentiality and meet Twitter policy, we do not share the names or usernames of the selected accounts.
We collected all elite users' tweets between January 1, 2022 and July 31, 2022. 
Based on relevant vaccine-related hashtags (that we stripped from the tweets) and a random tweet corpus~\cite{turenne2018rumour}, we trained a two-class classifier based on BertTweetFR~\cite{guo-etal-2021-bertweetfr}. This classifier allowed us to keep only elite users' tweets dealing with the vaccination debate. Here, we focus on retweets, which are signs of approval and thus give information about what users agree with~\cite{conover2011political}. Thus, we collected information about 100 randomly selected retweeters for each collected tweet, that we hope to be representative of all users. Among the selected retweeters, we focused on the 1,000 most active ones (500 pro-vaccine and 500 anti-vaccine).

\subsection{Data Analysis}

We collected 6,697 tweets in the period, divided into 1,869 tweets from pro-vaccine elite users, and 4,828 tweets from anti-vaccine elite users. From the 1,000 most active retweeters, we collected 11,449,936 retweets. 299,879 of these retweets were on elite users' tweets, with 16,791 retweets on pro-vaccine tweets, and 283,088 retweets on anti-vaccine tweets. This reflects a more intensive activity on the anti-vaccine side, which is consistent with the fact that anti-vaccine supporters are more engaged on Twitter, especially by doing many replies and retweets~\cite{germani2021anti}. Looking at the structure of the graph, we identify 2 highly connected sets of nodes (\textit{modularity}=0.55 \cite{clauset2004finding}), with few edges between them.
Besides, the controversy of the vaccine topic, computed using the Random Walk process described by Garimella~\textit{et al.}~\cite{garimella2018quantifying}, is equal to 0.89. This indicates that it is difficult to move from one community to the other one. Altogether, these results first confirm that the selected elite users are tweeting pro-vaccine and anti-vaccine according to the opinion for which they were chosen. Second, they corroborate polarized attitudes towards the vaccination debate within our dataset. 

\section{Towards a Multi-factorial analysis of Polarization Behaviors}
\label{multi_analysis}
In this section, we first analyze the information returned by each polarization metric separately, relying on a single factor. Second, we conduct a multi-factor analysis (bi-factor and tri-factor) by considering several metrics. We evaluate the ability to accurately differentiate and characterize polarization behaviors in these different experimental conditions.

\subsection{A Single Factor Analysis: Metrics from the Literature}
\label{1D}

To study polarization, we separately study two factors: (1) \textbf{Opinion factor}, where opinions are assessed from the standard users' retweets on each community (pro- or anti-vaccine); (2) \textbf{Source factor}, where sources are assessed from standard users' retweets on each elite user, who act as sources. To quantify polarization on these two factors, we rely on  two individual metrics: the \textit{polarization score} $\rho$~\cite{becatti2019extracting} and the \textit{Lack of Diversity} $LD$~\cite{cicchini2022news} (see Section~\ref{SOA}).  

Here, as we deal with single factor data, we use Kernel Density Estimation (KDE) rather than well-known multidimensional clustering algorithms to differentiate potential clusters. KDE estimates the probability density function of the studied single factor clusters based on local minima and maxima. 

Applied to $\rho$, the estimated kernel density has no local maxima or minima. This does not allow the differentiation of well-separated clusters. Looking closer at the values of $\rho$ for all standard users, we note that it is not well distributed in $[0, 1]$, and the minimum value is $0.5$. This is one of the limits of the metric (See Equation~(\ref{becatti_metric})), which is bounded in $[0.5, 1]$ in a 2-community context. Users having at least one interaction in each of the communities represent only 18.8\% of users, and KDE does not estimate a probability density function allowing to clearly differentiate them. Thus, $\rho$ cannot accurately and finely characterize behaviors depending on whether users interact with 1 or 2 communities. In the same way, the kernel density estimated on $LD$ does not have local minima or maxima as values are distributed continuously. Therefore, this metric does not allow for a differentiation of polarization behavior in terms of access to information sources neither.

\subsection{A Bi-factor Analysis}
\label{2D}

In this second step, we rely on a clustering algorithm to identify behavior clusters by combining opinion and source factors. We choose $k$-means~\cite{likas_global_2003} as it is well suited when dealing with numerical features. 
The number of clusters $k$ is optimized by maximizing 2 traditional metrics: Davies-Bouldin Index~\cite{davies1979cluster} (the lower the better) and Silhouette Index~\cite{rousseeuw_silhouettes_1987} (the higher the better).

\subsubsection{Use of Metrics from the Literature}
\label{2D-baselines}

The optimal number of clusters obtained with $\rho$ and $LD$ factors is $k=2$, where Davies-Bouldin Index $= 0.60$ and Silhouette index $=0.56$. Unlike the single factor analysis, considering the two metrics together does allow for the identification of two distinct groups of users, who adopt different polarization behaviors. This bi-factor analysis results in a finer-grained modeling. Identified clusters are shown in Figure~\ref{fig:clusters_baselines_2d}. Here again, $\rho$ does not allow a clear differentiation as users interacting with a unique community and those interacting in both communities are not associated with different clusters. Users are clustered according to the $LD$ value, with the two clusters being separated by a threshold fixed around $LD = 0.6$. Users in the orange cluster C1 with higher values of $LD$ are those with a high polarization according to accessed sources (\textit{i.e.} retweeted elite users), while the other users (blue cluster C2) retweet more elite users and in a balanced way. However, following the KDE estimation presented in Section \ref{1D}, this threshold is difficult to interpret. Although allowing for the differentiation of two groups of users, which was not possible with a single factor analysis, the limitations identified with respect to the distribution of $\rho$ values and the $LD$ threshold used to delimit the clusters question the quality of the clustering step. We expect the identified clusters to group together users that are likely to adopt well-differentiated polarization behaviors. 

\subsubsection{Refining Metrics from the Literature: Use of Entropy}
\label{2D-entropy}
 
As previously mentioned, from our analysis, assessing individual polarization by only considering the predominant opinion ($\rho$) or source ($LD$) limits the ability to differentiate between users, and to understand polarization behaviors. To address this limitation, we propose to consider all interactions and represent them as a probability distribution. Following Information Theory, this makes it possible to compute entropy~\cite{shannon1948mathematical}.  We thus propose to compute entropy-based metrics, measuring the uncertainty of access to opinions or sources. 

Precisely, the more homogeneously distributed the probability mass, the higher the entropy and the greater the uncertainty. 
As the maximal entropy depends on the number of entities, we propose to use the normalized entropy, $H_N(Z)= \frac{-\sum_{z}^n P(z)log(P(z))}{log(n)}$, 
where $Z$ is a discrete random variable that takes $n$ possible values and $P(z)$ is the probability of entity $z$. For simplicity's sake, from now on, we will refer to normalized entropy as entropy. As entropy is null when there is no randomness and equal to one when the behavior is very heterogeneous, in this work we use $H'(X) = 1-H_N(X)$ to get high scores for polarized users. To the best of our knowledge, no individual polarization metrics from the literature rely on it.

We note $H'_{op}$ the entropy-based opinion metric, and $H'_{so}$ the entropy-based source metric. First, comparing $\rho$ and $H'_{op}$, we can notice that unlike $\rho$, $H'_{op}$ ranges in $[0,1]$. Though, as a very large proportion of users only interact with a single community, their polarization on opinion remains maximal, with $\rho = H'_{op} = 1$. The potential contribution of entropy to the differentiation of polarization behaviors will therefore be for the other users ($H'_{op} \neq 1$), representing $18.8\%$ of standard users. Second, comparing $LD$ and $H'_{so}$, we notice that values are distributed differently. First of all, $LD$ values range in $[0.17, 1]$ while $H'_{so}$ values range in $[0.09,1]$. Second, the mean of $LD$ values is $0.59$, while the mean of $H'_{so}$ values is $0.50$. To go deeper into this comparison, we analyze the ranked values of $LD$ and $H'_{so}$, and get a Spearman score of 0.92 (\textit{p-value}$= 0$). Although relatively high, this score indicates that a significant proportion of users observe large rank variations. Actually, $49.5\%$ of users have a variation higher than $5\%$ of the positions. Given a user $u_1$, who has more than half of her interactions with one elite user, $LD(u_1) = 0.52$. Looking at $u_1$'s other retweets, they are on only 3 other elite users, with two of them having only one retweet. This is reflected in $H'_{so}(u_1) = 0.91$, which indicates a high level of polarization.  For a significant proportion of users, the information returned by $H'_{so}$ is well different from that returned by $LD$, as for $u_1$, and could potentially allow for the identification of different classes of polarization behaviors.

\begin{figure}[!ht]
\begin{subfigure}[t]{0.28\textwidth}
  \includegraphics[width=\linewidth]{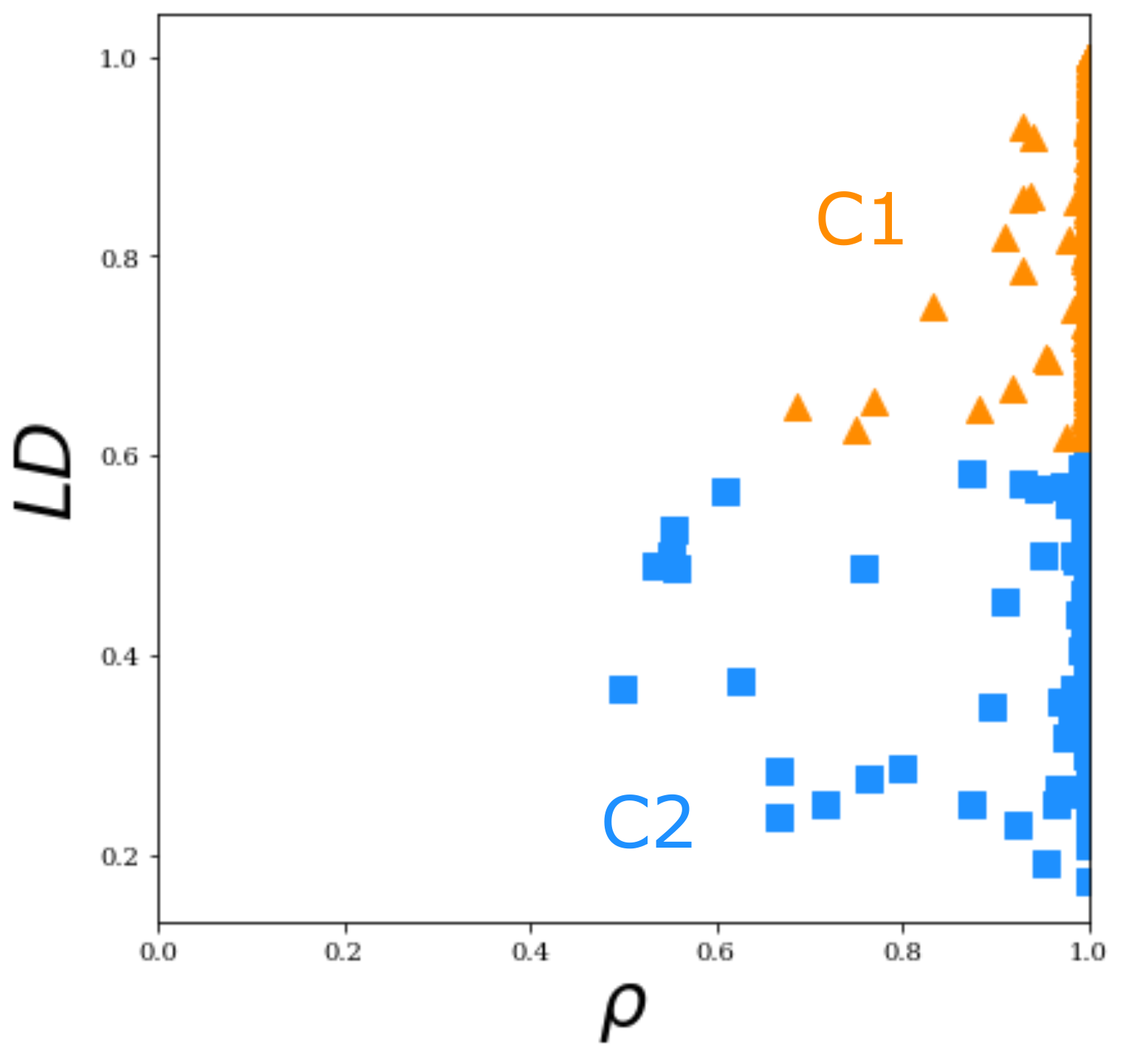}
  \caption{$\rho$ and LD.}
  \label{fig:clusters_baselines_2d}
\end{subfigure}
\begin{subfigure}[t]{0.28\textwidth}
  \includegraphics[width=\linewidth]{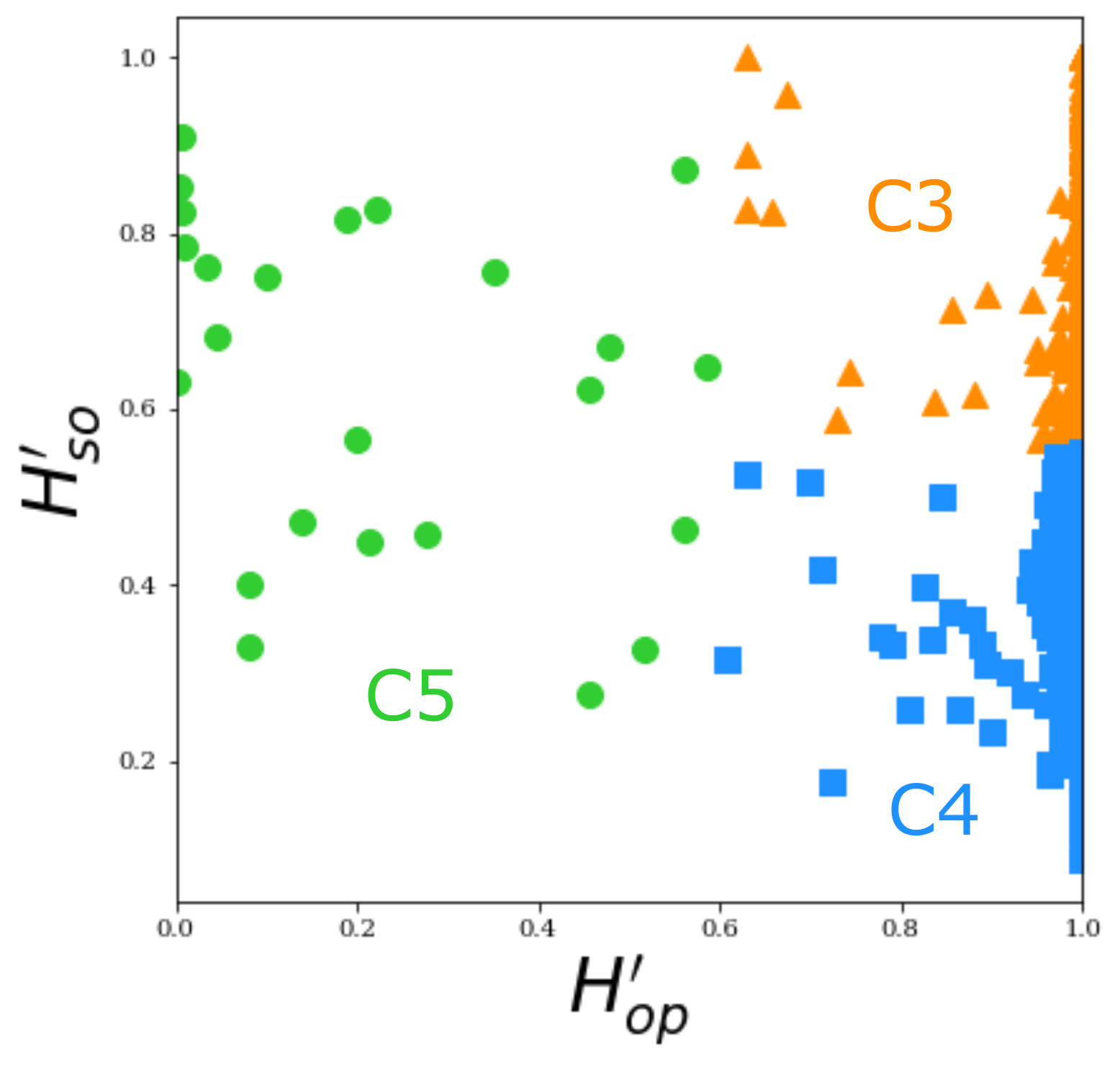}
  \caption{$H'_{op}$ and $H'_{so}$.}
  \label{fig:clusters_entropy_2d}
\end{subfigure}
\hspace{0.3cm}
\begin{subfigure}[t]{0.35\textwidth}
    \centering
    \includegraphics[width=1\linewidth]{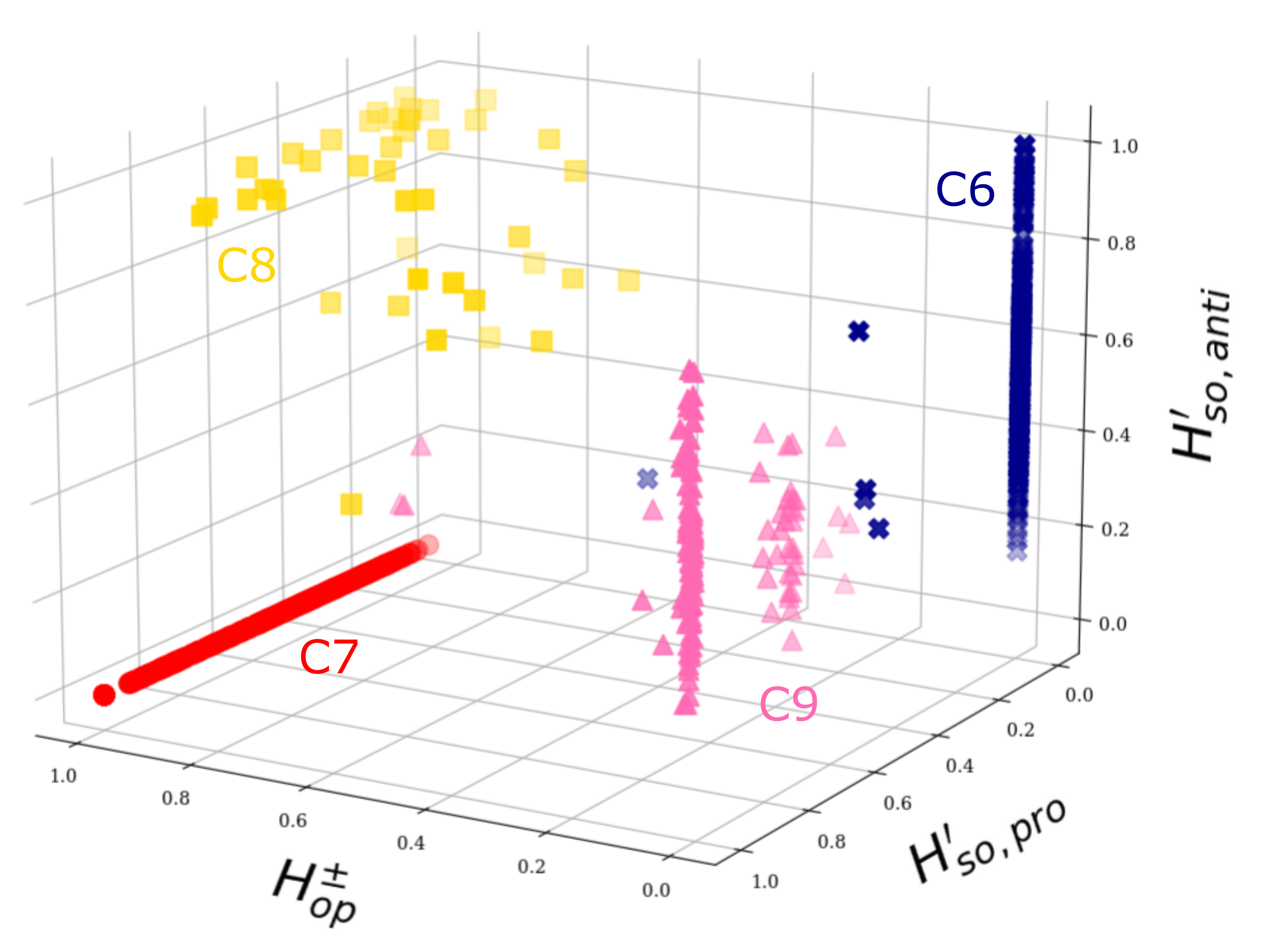}
    \caption{$H^{\pm}_{op}$, $H'_{so, pro}$ and  $H'_{so, anti}$}
    \label{fig:clusters_entropy_3d}
\end{subfigure}
\caption{Clusters identified with bi-factor data (3a and 3b) and tri-factor data (3c).}
\label{fig:clusters_2D}
\end{figure}

Applying the $k$-means algorithm on bi-factor data, namely $H'_{op}$ and $H'_{so}$, the optimal number of clusters is $k=3$, with Davies-Bouldin Index $=0.58$ and Silhouette Index $= 0.57$. Performance is thus close to the one with $\rho$ and $LD$, but one additional cluster is identified. Clusters are represented in  Figure~\ref{fig:clusters_entropy_2d}, with orange and blue clusters (C3 and C4) quite similar to the ones identified in Section~\ref{2D-baselines}.
The green cluster (C5), corresponds to a subset of 24 users with lower $H'_{op}$ values, thus interacting with both communities. In an unprecedented way, entropy-based metrics thus allow differentiating standard users both on opinion and source factors, which was not the case with $\rho$ and $LD$. More importantly, clustering on $H'_{op}$ and $H'_{so}$ contributes to identifying a new subset of users, interacting with both communities and potentially acting as intermediates between pro-vaccine and anti-vaccine users. The resulting clusters thus provide useful observations about the polarization of  users on SM. A bi-factor analysis, coupled with the use of entropy-based metrics, has made it possible to differentiate new classes of behavior. We now wish to assess to what extent  additional factors could further improve the quality of the polarization behaviors modeling.

\subsection{A Tri-factor Analysis}

One main limit of the literature and of the previous sections is that the metrics evaluate to what extent users are polarized, but do not inform towards which community. We assume that this could improve the clustering.

In this respect, and considering the opinion metric, we propose to apply a transformation factor, as follows: 
\begin{equation}
    \label{oriented_opinion_metric}
    H^{\pm}_{op} = \frac{\pm H'_{op} + 1}{2}
\end{equation}

The plus-minus sign in front of $H'_{op}$ depends on the predominant community. We set $H'_{op} > 0$ if interactions are in favor of the pro-vaccine community, and  $H'_{op} < 0$ otherwise. The final $H^{\pm}_{op}$ values range in $[0,1]$, with $H^{\pm}_{op}=0$ indicating a very high polarization in the anti-vaccine community, $H^{\pm}_{op}=1$, corresponding to an extreme polarisation in the pro-vaccine community and $H^{\pm}_{op} = 0.5$ reflecting balanced interactions between the two communities. 

Considering the source metric, and still to inform about each community,  we propose to split it into two metrics. The entropy-based metric $H'_{so}$, which initially combined sources from both communities, is split into two metrics $H'_{so, pro}$ and $H'_{so, anti}$, corresponding to the source factor computed on either pro or anti-vaccine community. This is designed to better differentiate users with unbalanced polarization in each of the two communities.

The optimization of $k$-means algorithm on the tri-factor data indicates that the optimal value of $k$ is 4, with Davies-Bouldin Index $= 0.51$ and Silhouette Index $= 0.74$. Performance is thus significantly higher than with two factors. 

Looking at Figure~\ref{fig:clusters_entropy_3d}, representing identified clusters, we see that the clusters identified are quite different from those described in Section~\ref{2D}. First, blue and red clusters (C6 and C7) respectively correspond to highly polarized anti-vaccine users and pro-vaccine users. These users, accessing a unique community, do not all behave the same way according to the source factor as values are distributed in $[0,1]$. 
Besides, the yellow and pink clusters (C8 and C9) are of particular interest. Looking more closely at the opinion and source metrics within these clusters, users in the yellow cluster (C8) are those having $H^{\pm}_{op}$ values close to $0.5$, indicating a balanced activity between the two communities. In each accessed community, these users interact with a variety of sources, as $H'_{so, pro}$ and $H'_{so, anti}$ values are evenly distributed among users. Besides, users associated with the pink cluster (C9) interact predominantly with the anti-vaccine community. Nevertheless, they all have at least 1 interaction in the pro-vaccine community, in which they mostly interact with a few elite users ($H'_{so, pro} \approx 1$). Users taking part in these two unprecedented clusters are intermediate users as they interact in both communities. Moreover, in addition to being quite different from those identified in Section~\ref{2D-entropy}, they are also much more numerous. Yellow and pink clusters (C8 and C9) contain 43 and 140 standard users respectively, while only 24 intermediate users were previously identified. It appears that a significant proportion of users do not engage in extreme polarizing behaviors. 

Overall, the identification of four patterns of polarization, only possible with the last analysis, is very interesting and reflects the multi-factorial nature of polarization. Identified behavioral classes are well-separated according to specific characteristics, which was not the case when considering only one or two factors on traditional or entropy-based metrics (Sections~\ref{1D} and~\ref{2D}). Nevertheless, a few users are at the intersection between clusters, whose membership to one or other of the groups is questionable. For example, some users associated with the blue cluster C6 (\textit{i.e.} highly polarized users in the anti-vaccine debate community) are also very close to the pink cluster C9. These users may be in a transitional phase which might be of interest.

\section{Conclusion and Perspectives}
\label{CCL}

The literature still lacks individual polarization metrics allowing fine-grained modeling of users' polarization behaviors. As the latter are the result of multiple influences, we conducted a multi-factorial analysis of polarization. 
Experiments confirm, first, that metrics from the literature using maximum value are too restrictive, and that the proposed entropy-based metrics allow a finer distinction between polarization behaviors, both for opinion and source factors. It does indeed contribute to the identification of an additional cluster of under-represented users, who have retweet interactions in both communities. These users adopt moderately polarized behavior about a highly polarized topic on SM. Altogether, these results indicate that current polarization metrics do not distinguish polarization behaviors properly (\textbf{RQ1}), and that entropy-based metrics seem better adapted.
Besides, conducting a tri-factor analysis allows an unprecedented identification of well-separated behavioral clusters, which emphasizes that an adequate combination of factors leads to more reliable modeling of polarization behaviors (\textbf{RQ2}). 

In a process of opening the filter bubbles, and reducing the polarization phenomenon, such a multi-factorial analysis could be greatly beneficial. In a strongly polarized context, within which users have formed strong opinions, an input of diverse items does not always have the desired effect. In fact, providing diversity can be tricky due to the strong opinions held by users, who are potentially very wary of being exposed to contrary ideas. It may even reinforce the polarization phenomenon \cite{bail2018opposing}. An accurate characterization of adopted polarization behaviors could help to adapt solutions, including through recommendations. Each identified behavioral class could benefit from different recommendation strategies. To go further, the intermediate classes of users identified through the multi-factorial analysis could help to gradually expose highly polarized users to different viewpoints. This could limit the potential adverse effects of diversity, and help build trust-based recommendations. Especially, users whose membership to one unique cluster is uncertain could serve as bridges between the different classes close to them. The future development of depolarizing recommender systems could probably rely on a multi-factorial analysis of polarization to limit this worrying phenomenon.

\bibliographystyle{unsrt}
\bibliography{biblio}

\begin{thebibliography}{10}

\bibitem{azzimonti2022polarization}
Marina Azzimonti and Marcos Fernandes.
\newblock Social media networks, fake news, and polarization.
\newblock {\em European Journal of Political Economy}, page 102256, 2022.

\bibitem{bail2018opposing}
Christopher~A. Bail, Lisa~P. Argyle, Taylor~W. Brown, John~P. Bumpus, Haohan
  Chen, M.~B.~Fallin Hunzaker, Jaemin Lee, Marcus Mann, Friedolin Merhout, and
  Alexander Volfovsky.
\newblock Exposure to opposing views on social media can increase political
  polarization.
\newblock {\em Proceedings of the National Academy of Sciences},
  115(37):9216--9221, 2018.

\bibitem{bernstein2020diversity}
Abraham Bernstein, Claes De~Vreese, Natali Helberger, Wolfgang Schulz,
  Katharina Zweig, Christian Baden, Michael~A Beam, Marc~P Hauer, Lucien Heitz,
  Pascal J{\"u}rgens, et~al.
\newblock Diversity in news recommendations.
\newblock {\em arXiv preprint arXiv:2005.09495}, 2020.

\bibitem{stray2021designingrs}
Jonathan Stray.
\newblock Designing recommender systems to depolarize.
\newblock {\em First Monday}, 27, 2021.

\bibitem{treuillier2022being}
Celina Treuillier, Sylvain Castagnos, Evan Dufraisse, and Armelle Brun.
\newblock Being diverse is not enough: Rethinking diversity evaluation to meet
  challenges of news recommender systems.
\newblock In {\em Fairness in User Modeling, Adaptation and Personalization
  (FairUMAP 2022)}, 2022.

\bibitem{helberger2018diversity}
Natali Helberger, Kari Karppinen, and Lucia D’Acunto.
\newblock Exposure diversity as a design principle for recommender systems.
\newblock {\em Information, Communication \& Society}, 21(2):191--207, 2018.

\bibitem{lunardi2020filter}
Gabriel~Machado Lunardi, Guilherme~Medeiros Machado, Vinicius Maran, and José
  Palazzo~M. {de Oliveira}.
\newblock A metric for filter bubble measurement in recommender algorithms
  considering the news domain.
\newblock {\em Applied Soft Computing}, 97:106771, 2020.

\bibitem{heitz2022benefits}
Lucien Heitz, Juliane~A Lischka, Alena Birrer, Bibek Paudel, Suzanne Tolmeijer,
  Laura Laugwitz, and Abraham Bernstein.
\newblock Benefits of diverse news recommendations for democracy: A user study.
\newblock {\em Digital Journalism}, pages 1--21, 2022.

\bibitem{badami2017peekingit}
Mahsa Badami.
\newblock {\em Peeking into the other half of the glass: handling polarization
  in recommender systems}.
\newblock PhD thesis, University of Louisville, 2017.

\bibitem{moller2018not}
Judith M{\"o}ller, Damian Trilling, Natali Helberger, and Bram van Es.
\newblock Do not blame it on the algorithm: an empirical assessment of multiple
  recommender systems and their impact on content diversity.
\newblock {\em Information, Communication \& Society}, 21(7):959--977, 2018.

\bibitem{kubin2021role}
Emily Kubin and Christian von Sikorski.
\newblock The role of (social) media in political polarization: a systematic
  review.
\newblock {\em Annals of the International Communication Association},
  45(3):188--206, 2021.

\bibitem{vanbavel2021media}
Jay~J. Van~Bavel, Steve Rathje, Elizabeth Harris, Claire Robertson, and Anni
  Sternisko.
\newblock How social media shapes polarization.
\newblock {\em Trends in Cognitive Sciences}, 25(11):913--916, 2021.

\bibitem{conover2011political}
Michael Conover, Jacob Ratkiewicz, Matthew Francisco, Bruno Gon{\c{c}}alves,
  Filippo Menczer, and Alessandro Flammini.
\newblock Political polarization on twitter.
\newblock In {\em Proceedings of the international aaai conference on web and
  social media}, volume~5, pages 89--96, 2011.

\bibitem{newman2006modularity}
Mark~EJ Newman.
\newblock Modularity and community structure in networks.
\newblock {\em Proceedings of the national academy of sciences},
  103(23):8577--8582, 2006.

\bibitem{garimella2018quantifying}
Kiran Garimella, Gianmarco De~Francisci Morales, Aristides Gionis, and Michael
  Mathioudakis.
\newblock Quantifying controversy on social media.
\newblock {\em ACM Transactions on Social Computing}, 1(1):1--27, 2018.

\bibitem{morales2015measuring}
Alfredo~Jose Morales, Javier Borondo, Juan~Carlos Losada, and Rosa~M Benito.
\newblock Measuring political polarization: Twitter shows the two sides of
  venezuela.
\newblock {\em Chaos: An Interdisciplinary Journal of Nonlinear Science},
  25(3):033114, 2015.

\bibitem{geschke2019triple}
Daniel Geschke, Jan Lorenz, and Peter Holtz.
\newblock The triple-filter bubble: Using agent-based modelling to test a
  meta-theoretical framework for the emergence of filter bubbles and echo
  chambers.
\newblock {\em British Journal of Social Psychology}, 58(1):129--149, 2019.

\bibitem{becatti2019extracting}
Carolina Becatti, Guido Caldarelli, Renaud Lambiotte, and Fabio Saracco.
\newblock Extracting significant signal of news consumption from social
  networks: the case of twitter in italian political elections.
\newblock {\em Palgrave Communications}, 5(1):1--16, 2019.

\bibitem{schmidt2018polarization}
Ana~Luc{\'\i}a Schmidt, Fabiana Zollo, Antonio Scala, Cornelia Betsch, and
  Walter Quattrociocchi.
\newblock Polarization of the vaccination debate on facebook.
\newblock {\em Vaccine}, 36(25):3606--3612, 2018.

\bibitem{cicchini2022news}
Tomas Cicchini, Sofia~Morena Del~Pozo, Enzo Tagliazucchi, and Pablo Balenzuela.
\newblock News sharing on twitter reveals emergent fragmentation of media
  agenda and persistent polarization.
\newblock {\em EPJ Data Science}, 11(1):48, 2022.

\bibitem{chen2019modeling}
Tinggui Chen, Qianqian Li, Jianjun Yang, Guodong Cong, and Gongfa Li.
\newblock Modeling of the public opinion polarization process with the
  considerations of individual heterogeneity and dynamic conformity.
\newblock {\em Mathematics}, 7(10):917, 2019.

\bibitem{sunstein1999law}
Cass~R Sunstein.
\newblock The law of group polarization.
\newblock {\em University of Chicago Law School, John M. Olin Law \& Economics
  Working Paper}, (91), 1999.

\bibitem{primario2017measuring}
Simonetta Primario, Dario Borrelli, Luca Iandoli, Giuseppe Zollo, and Carlo
  Lipizzi.
\newblock Measuring polarization in twitter enabled in online political
  conversation: The case of 2016 us presidential election.
\newblock In {\em 2017 IEEE international conference on information reuse and
  integration (IRI)}, pages 607--613. IEEE, 2017.

\bibitem{turenne2018rumour}
Nicolas Turenne.
\newblock The rumour spectrum.
\newblock {\em PloS one}, 13(1):e0189080, 2018.

\bibitem{guo-etal-2021-bertweetfr}
Yanzhu Guo, Virgile Rennard, Christos Xypolopoulos, and Michalis Vazirgiannis.
\newblock {BERT}weet{FR} : Domain adaptation of pre-trained language models for
  {F}rench tweets.
\newblock In {\em Proceedings of the Seventh Workshop on Noisy User-generated
  Text (W-NUT 2021)}, pages 445--450, Online, November 2021. Association for
  Computational Linguistics.

\bibitem{germani2021anti}
Federico Germani and Nikola Biller-Andorno.
\newblock The anti-vaccination infodemic on social media: A behavioral
  analysis.
\newblock {\em PloS one}, 16(3):e0247642, 2021.

\bibitem{clauset2004finding}
Aaron Clauset, Mark~EJ Newman, and Cristopher Moore.
\newblock Finding community structure in very large networks.
\newblock {\em Physical review E}, 70(6):066111, 2004.

\bibitem{likas_global_2003}
Aristidis Likas, Nikos Vlassis, and Jakob J.~Verbeek.
\newblock The global k-means clustering algorithm.
\newblock {\em Pattern Recognition}, pages 451--461, 2003.

\bibitem{davies1979cluster}
David~L Davies and Donald~W Bouldin.
\newblock A cluster separation measure.
\newblock {\em IEEE transactions on pattern analysis and machine intelligence},
  (2):224--227, 1979.

\bibitem{rousseeuw_silhouettes_1987}
Peter~J. Rousseeuw.
\newblock Silhouettes: A graphical aid to the interpretation and validation of
  cluster analysis.
\newblock {\em Journal of computational and applied mathematics}, pages 53--65,
  1987.

\bibitem{shannon1948mathematical}
Claude~Elwood Shannon.
\newblock A mathematical theory of communication.
\newblock {\em The Bell system technical journal}, 27(3):379--423, 1948.

\end{thebibliography}






\end{document}